\documentclass[twocolumn,amsmath,amssymb,longbibliography,superscriptaddress,groupedaddress,aps,pra,10pt]{revtex4-2}
\usepackage[dvipsnames]{xcolor}
\usepackage{multirow}
\usepackage[separate-uncertainty = true,multi-part-units=single]{siunitx}
\usepackage{graphicx}
\usepackage[english]{babel}
\usepackage[utf8]{inputenc} 
\usepackage{ctable} 
\usepackage{booktabs}
\usepackage{float}
\usepackage{wasysym}
\usepackage{color, colortbl}
\usepackage{german}
\usepackage[colorlinks=true,allcolors=blue]{hyperref}
\usepackage{doipubmed}

\newcommand{\Kr}[1]{$^{#1}$Kr}

\definecolor{Lightblue}{rgb}{0.85,0.85,0.95}
\definecolor{Lightred}{rgb}{0.95,0.9,0.9}

\begin{document}
	
	\title{Enhancement of the \Kr{81} and \Kr{85} count rates by optical pumping}
	\author{Z.-Y. Zhang}

	\author{F. Ritterbusch}
	\email{florian@ustc.edu.cn}

	\author{W.-K. Hu}
	\author{X.-Z. Dong}
	\author{C. Y. Gao}
	\author{W. Jiang}
	\author{S.-Y. Liu}

	\author{Z.-T. Lu}
	\email{ztlu@ustc.edu.cn}

	\author{J. S. Wang}
	\author{G.-M. Yang}

	\affiliation{\medskip Hefei National Laboratory for Physical Sciences at the Microscale, Center for Excellence in Quantum Information and Quantum Physics, Chinese Academy of Sciences, University of Science and Technology of China, 96 Jinzhai Road, Hefei 230026, China}

	\begin{abstract}
		We report an increase of up to 60\% on the count rates of the rare \Kr{81} and \Kr{85} isotopes in the Atom Trap Trace Analysis method by enhancing the production of metastable atoms in the discharge source. Additional atoms in the metastable $ 1s_5 $ level (Paschen notation) are obtained via optically pumping the $1s_4-2p_6$ transition at \SI{819}{nm}. By solving the master equation for the system, we identify this transition to be the most suitable one and can describe the measured increase in metastable population as a function of the 819-nm laser power. We calculate the previously unknown isotope shifts and hyperfine splittings of the $1s_4-2p_6$ transition in \Kr{81} and \Kr{85}, and verify the results with count rate measurements. The demonstrated count-rate increase enables a corresponding decrease in the required sample sizes for \Kr{81} and \Kr{85} dating, a significant improvement for applications such as dating of ocean water and deep ice cores.	
	\end{abstract}
	
	
	\maketitle
	
	\section{Introduction}
	
	\begin{figure*}
		\centering
		\noindent \includegraphics[width=17cm]{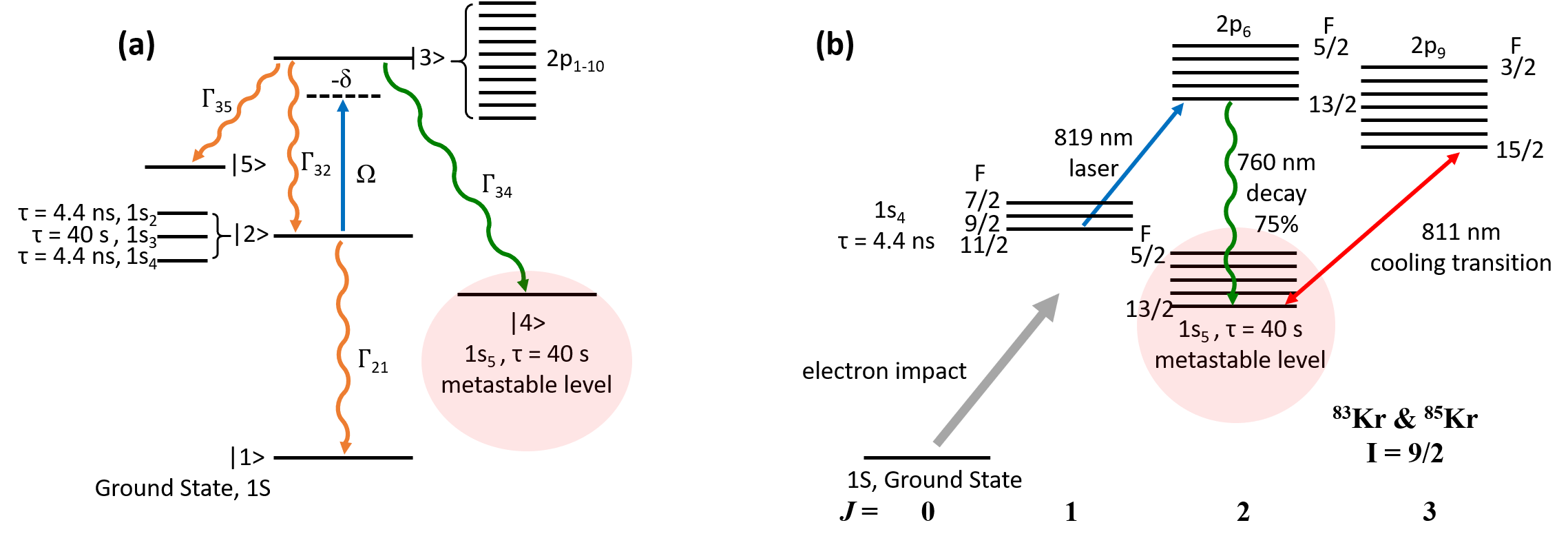}
		\caption{(a) Generic scheme for enhancing the population in the metastable state $ 1s_{5} $ by driving one of the $ 1s $ levels to one of the $ 2p $ levels with light of Rabi frequency $ \Omega $ and detuning $ \delta $. $ \Gamma _{ij} $ denotes the spontaneous emission rate from level $ i $ to $ j $.  
			(b) The optical pumping scheme chosen in this work on the $1s_4-2p_6$ transition at \SI{819}{nm}, shown for \Kr{83} and \Kr{85} (both have a nuclear spin $ I = 9/2 $). The schematic for \Kr{81} ($ I = 7/2 $) is similar.
		}
		\label{fig:scheme}
	\end{figure*}
	
	The noble gas radioisotopes \Kr{81} (half-life $t_{1/2}=\SI{229}{ka}$) and \Kr{85} ($t_{1/2}=\SI{11}{ a}$) are nearly ideal tracers for environmental processes owing to their chemical inertness and their gaseous properties \cite{Loosli1969, Lu2014}. However, these two tracers had been difficult to analyze due to their extremely low isotopic abundances in the range of $ 10^{-14}-10^{-11} $. In recent decades, the analytical method Atom Trap Trace Analysis (ATTA), which detects single atoms via their fluorescence in a magneto-optical trap, has made \Kr{81} dating available to the earth science community at large \cite{Chen1999, Jiang2012}. In the latest ATTA instrument, the required sample size for \Kr{81} dating has been reduced to \SI{1}{\mu L\ STP} of krypton, which can be extracted from $ 10-20 $ kg of water or ice \cite{Tian2019}. However, this sample requirement is still too large for several applications where \Kr{81} dating could help to resolve major questions in paleoclimatology, e.g. concerning the Greenland ice sheet stability or the Mid-Pleistocene Transition \cite{Schaefer2016, Yau2016, Severinghaus2010}.
	
	Laser cooling and trapping of krypton atoms in the ground level is not feasible due to the lack of suitable lasers at the required VUV wavelength. As is the case for all noble-gas elements, the krypton atoms first need to be excited to the metastable level $1s_5$ where the $1s_5-2p_9$ cycling transition at \SI{811}{nm} can be employed for laser cooling and trapping (Paschen notation \cite{Paschen1919} is used here, the corresponding levels in Racah notation \cite{Racah1942} can be found in Fig. \ref{transition} in Appendix \ref{krypton_transition_scheme_trans_frac}). The level $1s_5$ is \SI{10}{eV} above the ground level and, in operational ATTA instruments, is populated by electron-impact excitation in a RF-driven discharge with an efficiency of only $ 10^{-4}-10^{-3} $. By increasing this efficiency the count rate of \Kr{81} and \Kr{85} increases accordingly, resulting in a reduction of the needed sample size. 
	
	Since the discharge excites atoms into not only the metastable $1s_5$ but also several other excited levels, the metastable population can be enhanced by transferring atoms from these other excited levels to the metastable one via optical pumping (Fig. \ref{fig:scheme}). This mechanism has been demonstrated in a spectroscopy cell for argon with an increase of 81\% \cite{Hans2014,Frolian2015} and for xenon with an increase by a factor of 11 \cite{Hickman2016}. It has also been observed in a metastable beam of argon with an increase of 21\% \cite{Hans2014}. All these experiments were done on stable, abundant isotopes.
	
	In this work, we examine the enhancement of metastable production by optical pumping for both the stable, abundant and the radioactive, rare isotopes of krypton. To calculate the transfer efficiency to the metastable state as well as its power dependence, we solve the master equation for the corresponding multi-level system. Implementing the enhancement scheme for the rare \Kr{81} and \Kr{85} requires the respective frequency shifts for the $1s_4-2p_6$ transition at \SI{819}{nm}. We calculate these previously unknown isotope and hyperfine shifts and compare them to measurements of the \Kr{81} and \Kr{85} count rates as a function of frequency.
	
	\section{Theory}\label{sec:theory}
	
	The general transition scheme for enhancing the metastable production in a RF-driven discharge by optical pumping is illustrated in Fig. \ref{fig:scheme}. In the discharge, all excited levels are populated by electron-atom collisions, including the desired metastable level $ 1s_{5} $ and the other $ 1s $ levels. Additional atoms can be transferred into $ 1s_{5} $ by driving the transition from one of these $ 1s $ levels to a $ 2p $ level, followed by spontaneous decay. The enhancement depends on the population in the initial $ 1s $ level and the transfer efficiency. The transfer efficiency is calculated in the following for all possible $ 1s-2p $ transitions to identify suitable candidates.

	\subsection{Transfer efficiency}\label{theory_transfer_efficiency}
	We solve the Lindblad master equation (see details in Appendix \ref{Lindblad_master_equations}) for the 5-level system  shown in Fig. \ref{fig:scheme}(a) which corresponds to the even krypton isotopes without hyperfine structure. The atom is initially in level $ {\left |2 \right\rangle} $, i.e.,
	\begin{equation}
	\widetilde{\rho}_{22}(t=0)=1.
	\label{eq:1}
	\end{equation}
	The steady-state solution for the final population in the metastable state $ {\left |4 \right\rangle} $ then becomes
	\begin{gather}\label{eq:2}
		\widetilde{\rho}_{44}(t\to+\infty)=
		\\
		\frac{\Gamma _{34}}{\Gamma _{34}+\Gamma _{35}+\Gamma _{21}\{1+\frac{(\Gamma _{32}+\Gamma _{34}+\Gamma _{35})[(\Gamma _{21}+\Gamma _{32}+\Gamma _{34}+\Gamma _{35})^2+4\delta ^2]}{\Omega ^2(\Gamma _{21}+\Gamma _{32}+\Gamma _{34}+\Gamma _{35})}\}} \nonumber
	\end{gather}
	as a function of the laser detuning $ \delta $ and the Rabi frequency $\Omega$. In the case of unpolarized atoms, the Rabi frequency can be expressed as \cite{Steck2001}
	\begin{equation}
	\Omega ^2=\frac{2J_{3}+1}{2J_{2}+1}\frac{\lambda ^{3}\Gamma _{32}}{2\pi hc}I,
	\label{eq:3}
	\end{equation}
	where $ I $ is the intensity of the laser beam, $ \lambda=\SI{819}{nm} $, $ J_{2} $ and $ J_{3} $ are the angular momentum quantum numbers for level $ {\left |2 \right\rangle} $ and $ {\left |3 \right\rangle} $, respectively. In the resonant case ($ \delta =0 $) and infinite laser power, Eq. (\ref{eq:2}) simplifies to
	\begin{equation}
	\widetilde{\rho}_{44}(t\to+\infty)=\frac{\Gamma _{34}}{\Gamma _{34}+\Gamma _{35}+\Gamma _{21}},
	\label{eq:4}
	\end{equation}
	which is the maximum fraction that can be transferred to the metastable state $ {\left |4 \right\rangle} $. With Eqs. (\ref{eq:2}) and (\ref{eq:3}) we calculate the transfer efficiency $ \widetilde{\rho}_{44}(t\to+\infty) $ for the different $ 1s-2p $ transitions in \Kr{84} as a function of laser power. The transitions with the highest transfer efficiencies for the $ 1s $ levels are shown in Table \ref{tab:transfer_efficiencies} (see Table \ref{tab:trans_frac} in Appendix \ref{krypton_transition_scheme_trans_frac} for all transitions). 
	\begin{table}[H]
		\caption{\Kr{84} transitions with the highest transfer efficiencies from each $ 1s $ level, calculated for a laser beam with 9-mm diameter and different powers $ P $.}
		\centering
		\def\arraystretch{1.3} 
		\begin{tabular*}{\hsize}{@{}@{\extracolsep{\fill}}ccccc@{}}
			\hline\hline
			\multirow{2}{*}{\begin{tabular}[c]{@{}c@{}}Lower\\ level\end{tabular}} & \multirow{2}{*}{\begin{tabular}[c]{@{}c@{}}Upper\\ level\end{tabular}} & \multirow{2}{*}{$ \lambda \SI{}{(nm)}$} & \multicolumn{2}{c}{$ \widetilde{\rho}_{44}(t\to+\infty) $} \\ \cline{4-5} 
			&                              &                     & $ P=\SI{0.5}{W} $      & $ P\rightarrow +\infty \SI{}{W} $          \\ \hline
			$ 1s_{4} $                          & $ 2p_{6} $                          & 819              & 0.09       & 0.11      \\ 
			$ 1s_{3} $                          & $ 2p_{10} $                         & 1673             & 0.88       & 0.88      \\ 
			$ 1s_{2} $                          & $ 2p_{6} $                          & 1374             & 0.05       & 0.10      \\ \hline\hline
		\end{tabular*}
		\label{tab:transfer_efficiencies}
	\end{table}	
	From the metastable level $1s_3$ (see Fig. \ref{transition} in Appendix \ref{krypton_transition_scheme_trans_frac}), the $1s_3-2p_{10}$ transition at \SI{1673}{nm} has the highest transfer efficiency of almost 90\%. Since $1s_3$ is also metastable, only mW of laser power are needed to saturate the transition. However, experimentally we could only achieve an increase of the metastable \Kr{84} flux by about 10\% with this transition. This indicates that the population in the metastable $1s_3$ is only $ 10-20\% $ of that in the metastable $1s_5$.   The transfer efficiency from the $1s_2$ state is the highest for the 1374-nm transition to the $ 2p_{6} $ level. However, by measuring the output spectrum of the RF-driven discharge with a VUV monochrometer, we find that the $1s_2$ level is about four times less populated than the $ 1s_{4} $ level. We therefore conclude that the 819-nm transition from the $ 1s_{4} $ to the $ 2p_{6} $ level is the most promising candidate for increasing the metastable production in the krypton discharge. In the following we therefore focus on this transition as illustrated in Fig. \ref{fig:scheme}(b) for the odd isotopes \Kr{83} and \Kr{85}.  
	
	\subsection{Isotope shifts and hyperfine splittings for \Kr{81}, \Kr{83}, and \Kr{85}}\label{iso_hyper}
	\begin{figure*}[ht]
		\centering
		\noindent \includegraphics[width=17cm]{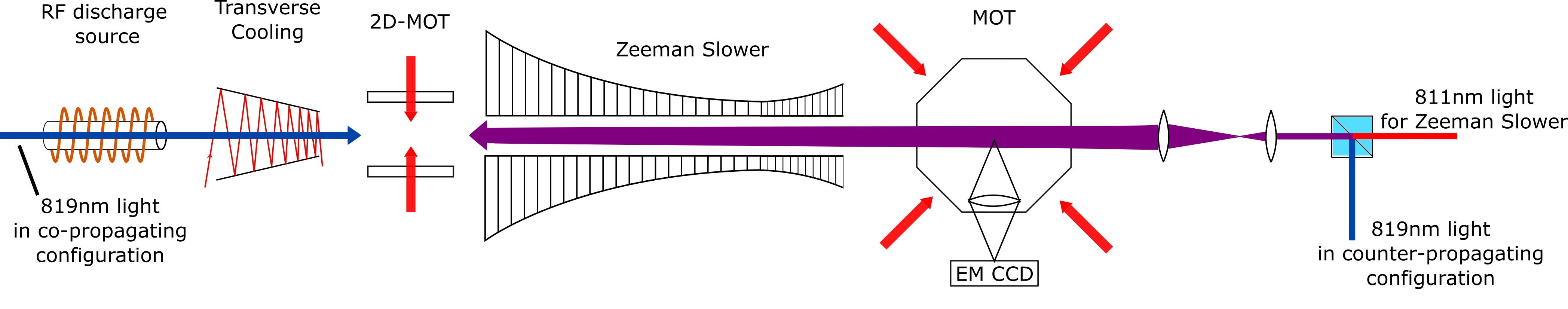}
		\caption{Sketch of the ATTA setup for investigating the enhancement of the \Kr{81} and \Kr{85} count rates by optical pumping.}
		\label{fig:setup}
	\end{figure*}
	For the 819-nm transition, the isotope shifts and hyperfine splittings of the odd krypton isotopes have not previously been measured. We therefore calculate approximate values based on measurements for other transitions and other isotopes as described in the following. 
	The hyperfine coefficients $ A $ and $ B $ depend on the specific isotope and energy level. For isotopes 1 and 2 they can in approximation be expressed as \cite{Armstrong1971}
	\begin{equation}
	\frac{A_1}{A_2}=\frac{\mu _{I}(1)}{\mu _{I}(2)}\frac{I(2)}{I(1)}
	\label{eq:5}
	\end{equation}and
	\begin{equation}
	\frac{B_1}{B_2}=\frac{Q(1)}{Q(2)},
	\label{eq:6}
	\end{equation}
	where $ \mu _{I} $  is the nuclear magnetic dipole moment and $ Q $ is the electric quadrupole moment. $ A $ and $ B $ for $ 2p_{6} $ of \Kr{81}, \Kr{83} and \Kr{85} were measured by Cannon \cite{Cannon1993}. $ A $ and $ B $ values for $ 1s_{4} $ of \Kr{83} were previously reported in \cite{Jackson1977} and $ \mu _{I} $ as well as $ Q $ were determined in \cite{Cannon1993}. With that, $ A $ and $ B $ for $ 1s_{4} $ of \Kr{81} and \Kr{85} can	be calculated using Eqs. (\ref{eq:5}) and (\ref{eq:6}). The relevant $ A $ and $ B $ values are listed in Table \ref{tab:A_B} in Appendix \ref{A_B_kr_cal}. With the resulting hyperfine constants for $ 1s_{4} $ and $ 2p_{6} $, the hyperfine shifts can be calculated.
	
	The isotope shifts of the even krypton isotopes for the 819-nm transition were measured in \cite{Jackson1979}. The unknown isotope shifts for the odd krypton isotopes can in first-order treatment be expressed as \cite{Heilig1974}
	\begin{equation}
	\begin{split}
	{\delta \nu _{i}^{X,X{}'}}&{=\nu _{i}^{X{}'}-\nu _{i}^{X}}
	\\
	&=F_{i}\delta \left \langle r^{2} \right \rangle^{X,X{}'}+\frac{X{}'-X}{XX{}'}M_{i},
	\label{eq:7}
	\end{split}
	\end{equation}
	where $ X,X{}' $ are atomic masses, $ F_{i} $ and $ M_{i} $ are coefficients of the 819-nm transition for all isotopes. Assuming that the mean square nuclear charge radius $ \delta \left \langle r^{2} \right \rangle^{X,X{}'} $ does not depend on the transition, $ F_{i} $ and $ M_{i} $ of the 819-nm transition can be calculated from the $ \delta \left \langle r^{2} \right \rangle^{X,X{}'} $ values determined previously on the 811-nm transition \cite{Keim1995}, along with the isotope shifts of the even krypton isotopes on the 819-nm transition \cite{Jackson1979}. The resulting isotope, hyperfine and total frequency shifts for the odd krypton isotopes relative to \Kr{84} are given in Table \ref{tab:kr_cal} in Appendix \ref{A_B_kr_cal}.

	\section{Experimental Setup}
	\begin{figure*}
		\centering
		\noindent \includegraphics[width=17cm]{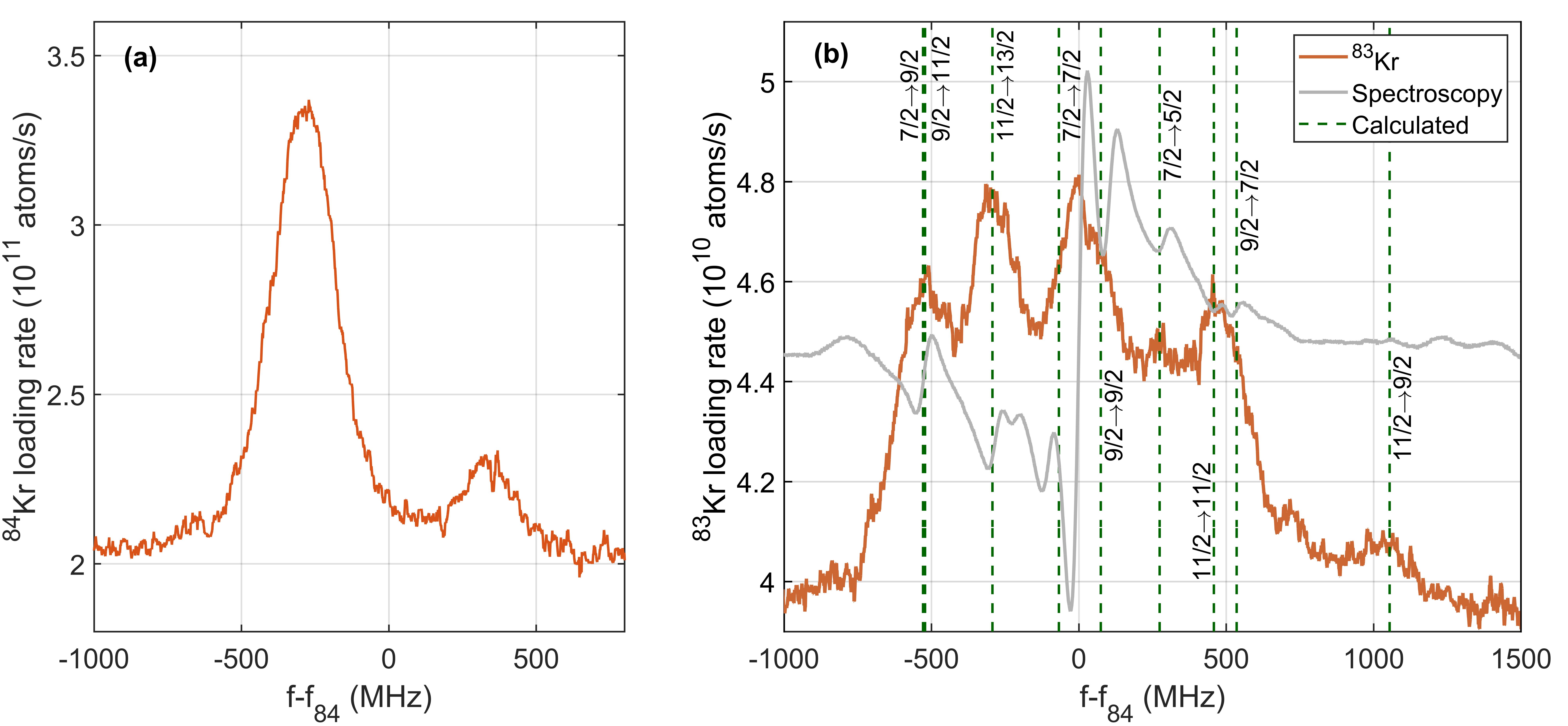}
		\caption{(a) \Kr{84} MOT loading rate versus detuning of the 819-nm laser. $ f_{84} $ denotes the resonance frequency of \Kr{84} at rest as monitored in a spectroscopy cell. The small peak on the positive frequency side is caused by a small fraction of the laser beam being reflected by the back window of the source.  (b) \Kr{83} MOT loading rate versus detuning of the 819-nm laser. The dashed lines mark the calculated resonances. The spectroscopy signal from a reference cell is plotted for comparison. The Doppler shift of the \Kr{83} loading-rate signal is removed to match the transitions as calculated and observed in the spectroscopy.}
		\label{fig:Kr84_Kr83}
	\end{figure*}
	An ATTA system (Fig. \ref{fig:setup}) is employed to measure the metastable enhancement by optical pumping on \Kr{81} and \Kr{85}. Metastable krypton atoms are produced in a RF-driven discharge via electron impact. The atomic beam emerging from the source is transversely cooled on the 811-nm transition in a tilted mirror arrangement. In the subsequent stage, the atomic beam is slightly focused in a two-dimensional magneto-optical trap (2D-MOT), and longitudinally slowed down in a Zeeman slower. The metastable atoms are then captured in the MOT, where single \Kr{81} and \Kr{85} atoms are detected via their 811-nm fluorescence using an EMCCD camera. The MOT loading rate of the abundant \Kr{83} (\Kr{83}/Kr $ \sim $11\%) is measured by depopulating the MOT with the quenching transition and detecting the emitted fluorescence \cite{Jiang2012}. The loading rate of the abundant \Kr{84} (\Kr{84}/Kr $ \sim $57\%) is determined by first clearing the MOT with the quenching transition and  then measuring the initial linear part of the rising slope of the MOT fluorescence \cite{Cheng2013}.
	
	Without applying optical pumping, the atom count rate is $ \sim\SI{10000}h^{-1} $ for \Kr{85} and $ \sim\SI{500}h^{-1} $ for \Kr{81} (lower than in \cite{Jiang2012} because the source is not liquid nitrogen cooled). After a measurement, the obtained \Kr{85} and \Kr{81} count rates are normalized by the \Kr{83} loading rate to account for drifts in the performance of the system.
	
	For optical pumping, we shine in the 819-nm light in two different configurations: (1) counter-propagating and (2) co-propagating to the atomic beam (Fig. \ref{fig:setup}). In the counter-propagating configuration, the 819-nm light is overlapped with the Zeeman slower beam and gently focused onto the exit of the source tube. In the co-propagating configuration, the 819-nm light comes in from the upstream side of the discharge source. In both configurations, the 819-nm light is delivered to the setup via a single-mode fiber. The available 819-nm power out of the fiber is around \SI{620}{mW}, which is generated by a tapered amplifier seeded by light from a diode-laser.  The 819-nm transition in the abundant krypton isotopes is monitored via modulation transfer spectroscopy \cite{Ma1990} in a reference cell. For measurements of \Kr{83}, \Kr{81}, and \Kr{85} in the ATTA setup, the frequency of the 819-nm laser needs to be tuned and stabilized over several hundred MHz. For that purpose, we use a scanning transfer cavity lock (STCL) \cite{Zhao1998, Subhankar2019}, using a diode laser locked to the 811-nm cooling transition of metastable \Kr{84} as a master laser.

	\section{Results and Discussion}
	\begin{figure*}[t]
		\centering
		\noindent \includegraphics[width=17cm]{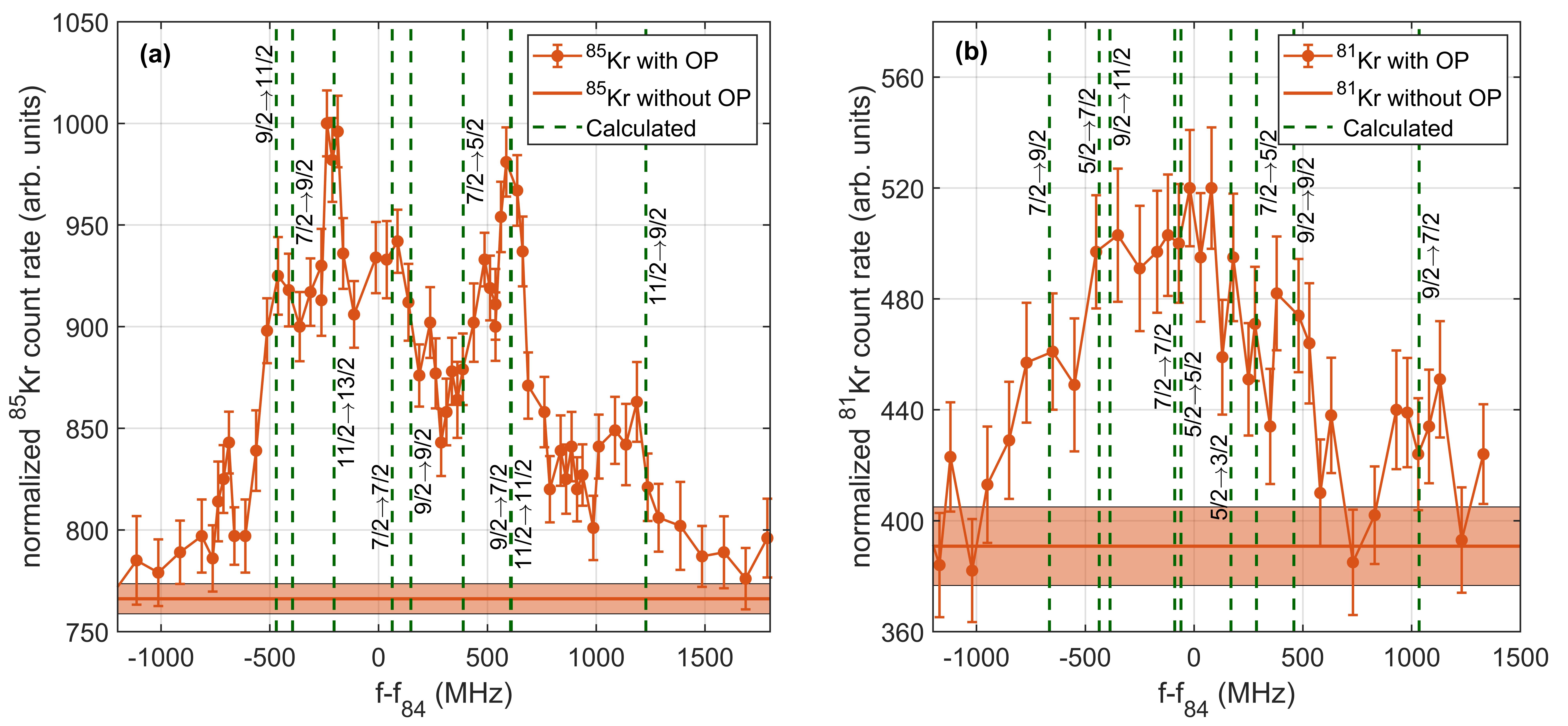}
		\caption{(a) Normalized \Kr{85} count rate versus frequency of the 819-nm laser, with and without optical pumping (OP). (b) Normalized \Kr{81} count rate versus frequency. The count rates are normalized by the \Kr{83} loading rate with the optical pumping light being resonant to the $ F=11/2\rightarrow13/2 $ transition.}
		\label{fig:Kr85_Kr81}
	\end{figure*}
	
	We measured the loading rates of different krypton isotopes versus frequency of the 819-nm laser. These measurements were done in counter-propagating configuration (Fig. \ref{fig:setup}). The MOT loading rate of the stable and abundant \Kr{84} versus the frequency of the 819-nm laser is shown in Fig. \ref{fig:Kr84_Kr83}(a). The most probable Doppler shift (\SI{-276}{MHz}) is significantly smaller than that of a thermal beam at room temperature (\SI{-364}{MHz}), likely because transverse cooling is more efficient for slower atoms. 
	
	The data for the odd isotope \Kr{83}, displaying its hyperfine structure, are given in Fig. \ref{fig:Kr84_Kr83}(b) together with the spectroscopy signal from a reference cell. The \Kr{83} loading-rate signal is frequency shifted to match the \Kr{83} transitions as calculated and observed in the reference cell. The resulting most probable Doppler shift is \SI{-240}{MHz} which is different from \Kr{84} by \SI{36}{MHz}. This difference is likely caused by the transverse cooling which is less efficient for \Kr{83} because of its hyperfine structure leading to slower atoms being collimated more efficiently. The largest increase is observed on the transitions $ F=11/2\rightarrow13/2 $ at \SI{-310}{MHz} and on the overlap of $ F=7/2\rightarrow7/2 $ and $ F=9/2\rightarrow9/2 $ at around \SI{0}{MHz}. To cover all three hyperfine states of the $ 1s_{4} $ state, we add sidebands to the laser light with an electro-optical modulator. When resonant with the $ F=11/2\rightarrow13/2 $ transition, we find the biggest increase for a sideband frequency of around \SI{240}{MHz}, likely because the $ F=7/2\rightarrow9/2 $ and $ F=9/2\rightarrow11/2 $ transitions are both addressed at this frequency. However, the resulting additional increase is only about 5\%. It seems that a significant fraction of the atoms in $ F=7/2 $ and $ F=9/2 $ is already addressed by the light resonant with the $ F=11/2\rightarrow13/2 $ transition due to power broadening. It may also be that the increase due to the sidebands is partly compensated by the loss of power at the central frequency.
	
	We applied this enhancement scheme to counting the extremely rare isotopes of krypton, \Kr{81} (isotopic abundance $ \sim1\times10^{-12} $) and \Kr{85} (isotopic abundance $ \sim2\times10^{-11} $). Fig. \ref{fig:Kr85_Kr81}(a) shows the \Kr{85} count rate normalized by the \Kr{83} loading rate versus the frequency of the 819-nm laser. We subtract the most probable Doppler shift of \SI{-240}{MHz} obtained from the \Kr{83} spectrum (assuming that the velocity distribution after transverse cooling is the same for the odd krypton isotopes) to obtain the frequency shifts at rest. The $ F=11/2\rightarrow13/2 $ transition displays the strongest increase as expected from the theoretical transition strength \cite{Axner2004}. Within the low level of statistics, the measured transition frequencies agree with the calculated ones except for the $ F=7/2\rightarrow5/2 $ transition where no distinct feature is observed.
	
	Fig. \ref{fig:Kr85_Kr81}(b) shows the \Kr{81} count rate normalized by the \Kr{83} loading rate versus the frequency of the 819-nm laser. As for \Kr{85}, we correct with the Doppler shift according to the \Kr{83} spectrum. The transition frequencies are less resolved than \Kr{85} due to the lower counting statistics. However, the peak corresponding to the $ F=9/2\rightarrow7/2 $ transition is clearly visible and, within the relatively large uncertainties, in agreement with the calculated frequency. The other hyperfine transitions cannot be resolved but they qualitatively agree with the calculation in that together they form a broad feature that contains the calculated transitions. 
	
	The measurements have so far been done in the counter-propagating configuration illustrated in Fig. \ref{fig:setup}. In this configuration the total 819-nm laser power is not used efficiently since the laser beam size is significantly larger than the inner diameter of the source tube. To increase the laser intensity we shine in the 819-nm light in co-propagating configuration (Fig. \ref{fig:setup}) with a laser beam size comparable to the inner diameter of the source tube. With the maximum available 819-nm power of \SI{620}{mW}, we obtain the loading rate increases shown in Table \ref{tab:loading_rate}. The increase for \Kr{84} is significantly higher than for \Kr{83}, presumably due to the insufficient coverage of the hyperfine levels of \Kr{83}. Within the uncertainties, the enhancements for the odd krypton isotopes are comparable.
	\begin{table}[h]
		\caption{Loading rate increases for the different krypton isotopes. The \Kr{85} and \Kr{81} count rates are normalized with the loading rate of \Kr{83}.\\}
		\centering
		\def\arraystretch{1.3} 
		\begin{tabular*}{\hsize}{@{}@{\extracolsep{\fill}}cccc@{}}
			\hline\hline
			Isotope                                                           & no OP     & with OP   & Increase                                           \\ \hline
			\begin{tabular}[c]{@{}l@{}}\Kr{84} {(}\SI{e11}{atoms/s}{)}\end{tabular} & 1.54(5)   & 2.82(10)  & \begin{tabular}[c]{@{}l@{}}83(9)\%\end{tabular} \\ 
			\begin{tabular}[c]{@{}l@{}}\Kr{83} {(}\SI{e10}{atoms/s}{)}\end{tabular} & 3.62(10)  & 5.38(16)  & 49(6)\%                                            \\ 
			\Kr{85} {(}arb. units{)}                                             & 629(17) & 986(29) & 57(6)\%                                            \\ 
			\Kr{81} {(}arb. units{)}                                              & 287(14)   & 476(13)   & 66(9)\%                                            \\ \hline\hline
		\end{tabular*}
		\label{tab:loading_rate}
	\end{table}
	\begin{figure}[h]
		\centering
		\noindent \includegraphics[width=9cm]{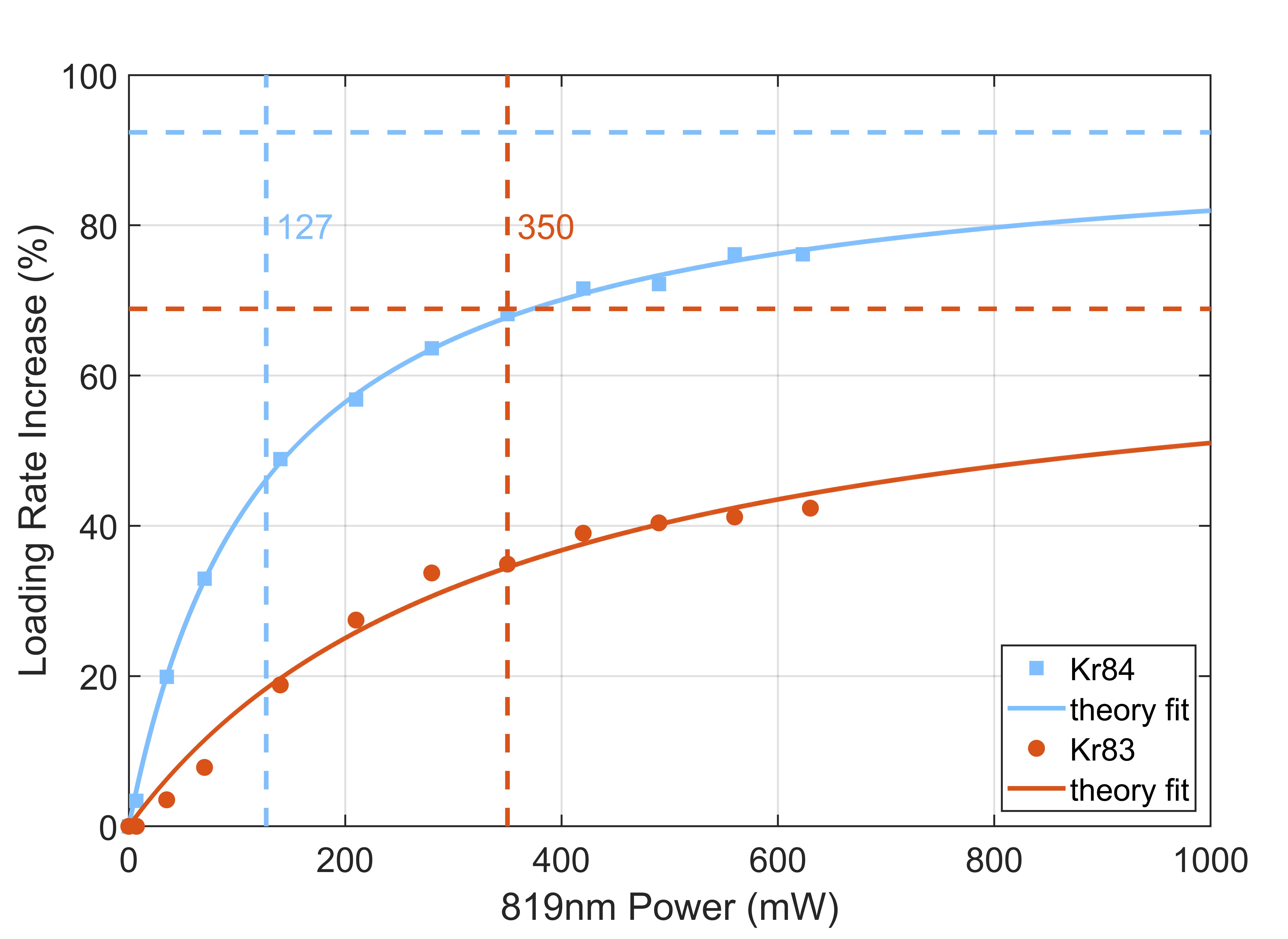}
		\caption{\Kr{83} (orange circles) and \Kr{84} (light blue squares) loading rate increase versus 819-nm laser power. The theory curves used for fitting are deduced in Sec. \ref{sec:theory}. They converge to the horizontal dashed lines for the case of infinite laser power. The vertical dashed lines indicate the power where half of the asymptotic value is reached.}
		\label{fig:power_fit}
	\end{figure}
	\newpage
	The loading rate increases of \Kr{84} and \Kr{83} as a function of the 819-nm laser power is shown in Fig. \ref{fig:power_fit}.
	We use Eqs. (\ref{eq:2}) and (\ref{eq:3}) to fit the power dependence leaving the amplitude and a scaling factor for the Rabi frequency (i.e. replacing $\Omega$ with $c\Omega$) as fit parameters. Since for \Kr{83} the closed $ F=11/2\rightarrow13/2 $ transition is employed, Eq. (\ref{eq:3}) can also be used by replacing the angular momentum numbers $ J $ with the corresponding total angular momentum numbers $ F $. We use the mean intensity over the cross-section of the source tube (9-mm diameter) for the 819-nm Gaussian beam (7.5-mm waist). From the fit for \Kr{84} we obtain a scaling factor for the Rabi frequency of 0.8. That the resulting scaling factor differs from unity might be due to omitting the velocity distribution in Eqs. (\ref{eq:2}) and (\ref{eq:3}), i.e. higher powers are needed to also address the off-resonant atoms. Moreover, the laser power is measured in front of the source window so it is likely lower at the interaction region due to transmission losses through the window and resonant krypton atoms in the discharge. For \Kr{83} the scaling factor for the Rabi frequency obtained from the fit is 0.6. This leads to a saturation power (where half of the asymptotic value is reached) 2.8 times higher than for \Kr{84}. This is likely caused by the hyperfine structure of \Kr{83}, which requires higher power to also saturate the sideband transitions. For infinite laser power the fits converge to an increase of 92\% and 69\% for \Kr{84} and \Kr{83}, respectively. According to Eq. (\ref{eq:4}) these values should be the same given that the spontaneous decay rates are the same for \Kr{84} and \Kr{83}. The discrepancy might be due to the large uncertainty in the fit especially for \Kr{83} or due to that the hyperfine structure of \Kr{83} is not accounted for in Eq. (\ref{eq:2}).
	
	\section{Conclusion and Outlook}
	We examined the use of optical pumping to enhance the production of metastable krypton in a RF-driven discharge source in an ATTA system. For \Kr{84}, at the maximum available laser power of \SI{620}{mW}, we reach an increase of 83\%, which is not far from the asymptotic value of 92\% at infinite laser power as extrapolated from the measured power dependence. For the rare \Kr{81} and \Kr{85} we obtain an increase of $ \sim $60\%, lower than for \Kr{84}, likely due to their hyperfine structures, which require higher laser power to also saturate the sideband transitions.
	
	The obtained results for the so far unknown frequency shifts of \Kr{81} and \Kr{85} at the 819-nm transition are an important contribution to the ongoing efforts to optically produce metastable \Kr{81} and \Kr{85} via resonant two-photon excitation \cite{Young2002,Ding2007,Daerr2011,Kohler2014}. For a precise measurement of the hyperfine coefficients and the isotope shifts, samples enriched in \Kr{81} and \Kr{85} will be necessary. 
	
	The results of this work enable the implementation of the presented method for enhanced production of metastable krypton in existing ATTA setups. This way, the sample size for \Kr{81} and \Kr{85} analysis in environmental samples can be significantly reduced, an improvement particularly important for radiokrypton dating of ocean water and deep ice cores.

\newpage
	
	\begin{acknowledgments}
		This work is funded by the National Key Research and Development Program of China (Grant No. 2016YFA0302200), National Natural Science Foundation of China  (Grant No. 41727901, No. 41861224007, and No. 11705196), and Anhui Initiative in Quantum Information Technologies (Grant No. AHY110000).
		
		\parskip=3mm
		
		\textit{An edited version of this paper was published by APS \href {https://doi.org/10.1103/PhysRevA.101.053429}{Physical Review A 101, 053429 (2020)}. Copyright 2020 American Physical Society.}
	\end{acknowledgments}

	\clearpage
	
	\appendix
	
	%
	

	\section{Master equation}\label{Lindblad_master_equations}
\begin{figure}[H]
		\centering
		\includegraphics[scale=0.5]{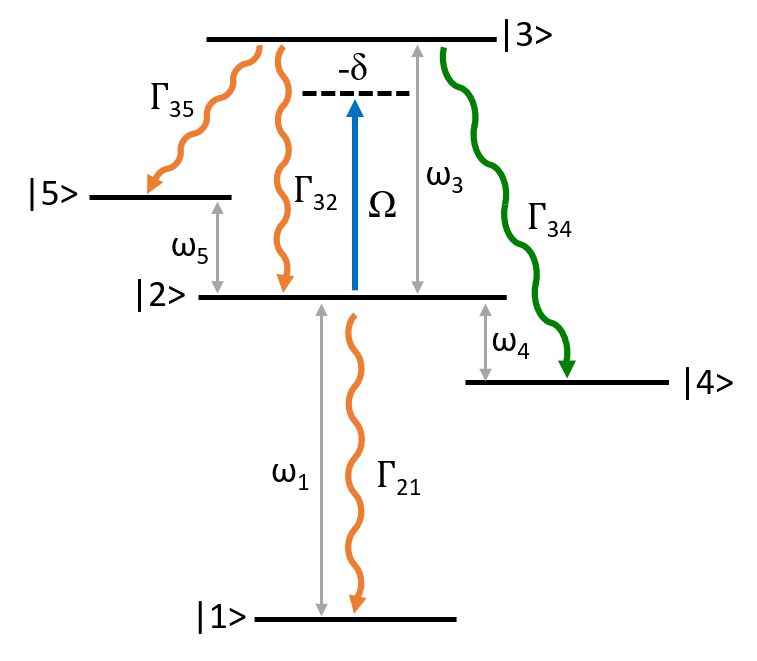}
		\caption{Five-level system describing the enhancement in the metastable level $\vert4\rangle$ by optical pumping. $ \Gamma _{ij} $ denotes the spontaneous emission rate from level $ i $ to $ j $, $\Omega$ the Rabi frequency and $\delta$ the detuning of the incident light with respect to the resonance frequency $\omega_3$.}
		\label{fig:scheme_appendix}
	\end{figure}

	The 5-level system for even isotopes without hyperfine structures which the metastable enhancement by optical optical pumping is based on, is illustrated in Fig. \ref{fig:scheme_appendix}. As in Fig. \ref{fig:scheme} in the introduction, $\vert1\rangle$ is the ground state and $\vert4\rangle$ the metastable state for laser cooling and trapping. Atoms in level $\vert2\rangle$ can be transferred to the metastable level $\vert4\rangle$ by driving the transition to level $\vert3\rangle$ followed by spontaneous decay. $\vert5\rangle$ represents other states that atoms can decay to from $\vert3\rangle$.
	Choosing the energy of level $\vert2\rangle$ as zero, $\hbar\omega_3,-\hbar\omega_1,-\hbar\omega_4$ and $\hbar\omega_5$ are the energies of the corresponding states relative to $\vert2\rangle$.
	In the Schr{\"o}dinger picture, the Hamiltonian of this atomic system interacting with the laser field is
	
	\begin{equation}
	\hat{H}=\hat{H}_A+\hat{H}_{AF},
	\end{equation} 
	where 
	\begin{equation}
	\hat{H}_A=\hbar\omega_3\vert 3\rangle\langle 3\vert-\hbar\omega_1\vert 1\rangle\langle 1\vert-\hbar\omega_4\vert 4\rangle\langle 4\vert+\hbar\omega_5\vert 5\rangle\langle 5\vert
	\end{equation}
	is the atomic Hamiltonian and 
	\begin{equation}
	\hat{H}_{AF}=\frac{\hbar\Omega}2(\sigma_{23}e^{i\omega t}+\sigma_{32}e^{-i\omega t}),
	\end{equation} 
	is the Hamiltonian that describes the interaction of the atoms with the light field. Here, $\Omega$ is the Rabi frequency of the incident light and $\sigma_{ij}=\vert i\rangle\langle j\vert$. 
	With the unitary transformation $U=\mathrm{exp}(i\omega t\vert 3\rangle\langle 3\vert)$, the quantum state $\vert\psi\rangle$ changes to
	\begin{equation}
	\widetilde{\vert\psi\rangle}=U\vert\psi\rangle.
	\end{equation}
	
	In this interaction picture, the Hamiltonian becomes
	\begin{equation}
	\begin{split}
	\widetilde{H}&=UHU^\dagger+i\hbar(\partial_tU)U^\dagger\\
	&=-\hbar\delta\sigma_{33}-\hbar\omega_1\sigma_{11}-\hbar\omega_4\sigma_{44}+\hbar\omega_5\sigma_{55}\\
	&\quad+\frac{\hbar\Omega}2(\sigma_{23}+\sigma_{32}),
	\end{split}
	\end{equation}
	where $\delta=\omega-\omega_3$ is the detuning of the light with respect to the transition frequency $\omega_3$ from $\vert2\rangle$ to $\vert3\rangle$.
	\par
	The Lindblad master equation for the system including the spontaneous emission can be written as
	\begin{equation}
	\begin{split}
	\frac{d\widetilde{\rho}}{dt}
	&=\frac{1}{i\hbar}[\widetilde{H}, \widetilde{\rho}]\\
	&\quad+\Gamma_{32}(\sigma_{23}\widetilde{\rho}\sigma_{32}-\frac12\widetilde{\rho}\sigma_{33}-\frac12\sigma_{33}\widetilde{\rho})\\
	&\quad+\Gamma_{21}(\sigma_{12}\widetilde{\rho}\sigma_{21}-\frac12\widetilde{\rho}\sigma_{22}-\frac12\sigma_{22}\widetilde{\rho})\\
	&\quad+\Gamma_{34}(\sigma_{43}\widetilde{\rho}\sigma_{34}-\frac12\widetilde{\rho}\sigma_{33}-\frac12\sigma_{33}\widetilde{\rho})\\
	&\quad+\Gamma_{35}(\sigma_{53}\widetilde{\rho}\sigma_{35}-\frac12\widetilde{\rho}\sigma_{33}-\frac12\sigma_{33}\widetilde{\rho}),
	\end{split}
	\end{equation}
	where $\Gamma_{ij}$ is the spontaneous emission rate from $\vert i\rangle$ to $\vert j\rangle$. These equations describe the time evolution of $\widetilde{\rho}_{ij}=\langle i\vert\widetilde{\rho}\vert j\rangle$ and can be simplified to
	\begin{equation}
	\begin{split}
	\frac{d\widetilde{\rho}_{11}}{dt}&=\Gamma_{21}\widetilde{\rho}_{22},\\
	\frac{d\widetilde{\rho}_{22}}{dt}&=\Gamma_{32}\widetilde{\rho}_{33}-\Gamma_{21}\widetilde{\rho}_{22}+\frac{i\Omega}{2}(\widetilde{\rho}_{23}-\widetilde{\rho}_{32}),\\
	\frac{d\widetilde{\rho}_{33}}{dt}&=-(\Gamma_{32}+\Gamma_{34}+\Gamma_{35})\widetilde{\rho}_{33}+\frac{i\Omega}{2}(\widetilde{\rho}_{32}-\widetilde{\rho}_{23}),\\
	\frac{d\widetilde{\rho}_{44}}{dt}&=\Gamma_{34}\widetilde{\rho}_{33},\\
	\frac{d\widetilde{\rho}_{55}}{dt}&=\Gamma_{35}\widetilde{\rho}_{33},\\
	\frac{d\widetilde{\rho}_{32}}{dt}&=-\frac12(\Gamma_{32}+\Gamma_{21}+\Gamma_{34}+\Gamma_{35}-2i\delta)\widetilde{\rho}_{32}\\
	&\quad+\frac{i\Omega}{2}(\widetilde{\rho}_{33}-\widetilde{\rho}_{22}).
	\label{eq:Lindblad}
	\end{split}
	\end{equation} 
	
	Using that the atom is initially in level $\vert2\rangle$, i.e. 
	\begin{equation}
	\begin{split}
	\widetilde{\rho}_{22}(t=0) &= 1   \\
	\widetilde{\rho}_{ij}(t=0)&=0 , \text{ }(i,j)\not=(2,2) , 
	\end{split}
	\end{equation}
	then for the steady-state
	\begin{equation}
	\frac{d\widetilde{\rho}}{dt}(t\rightarrow+\infty)=0
	\end{equation}
	Eq. \ref{eq:Lindblad} can be solved analytically, yielding the transfer efficiency $ \widetilde{\rho}_{44}(t\to+\infty) $ of Eq. \ref{eq:2}.
	
	\clearpage
	
	\onecolumngrid
	
		\section{TRANSFER EFFICIENCIES FOR $ 1s–2p $ TRANSITIONS IN \Kr{84}}\label{krypton_transition_scheme_trans_frac}
		
\begin{figure}[h]
			\centering
			\noindent \includegraphics[width=22cm,angle=270]{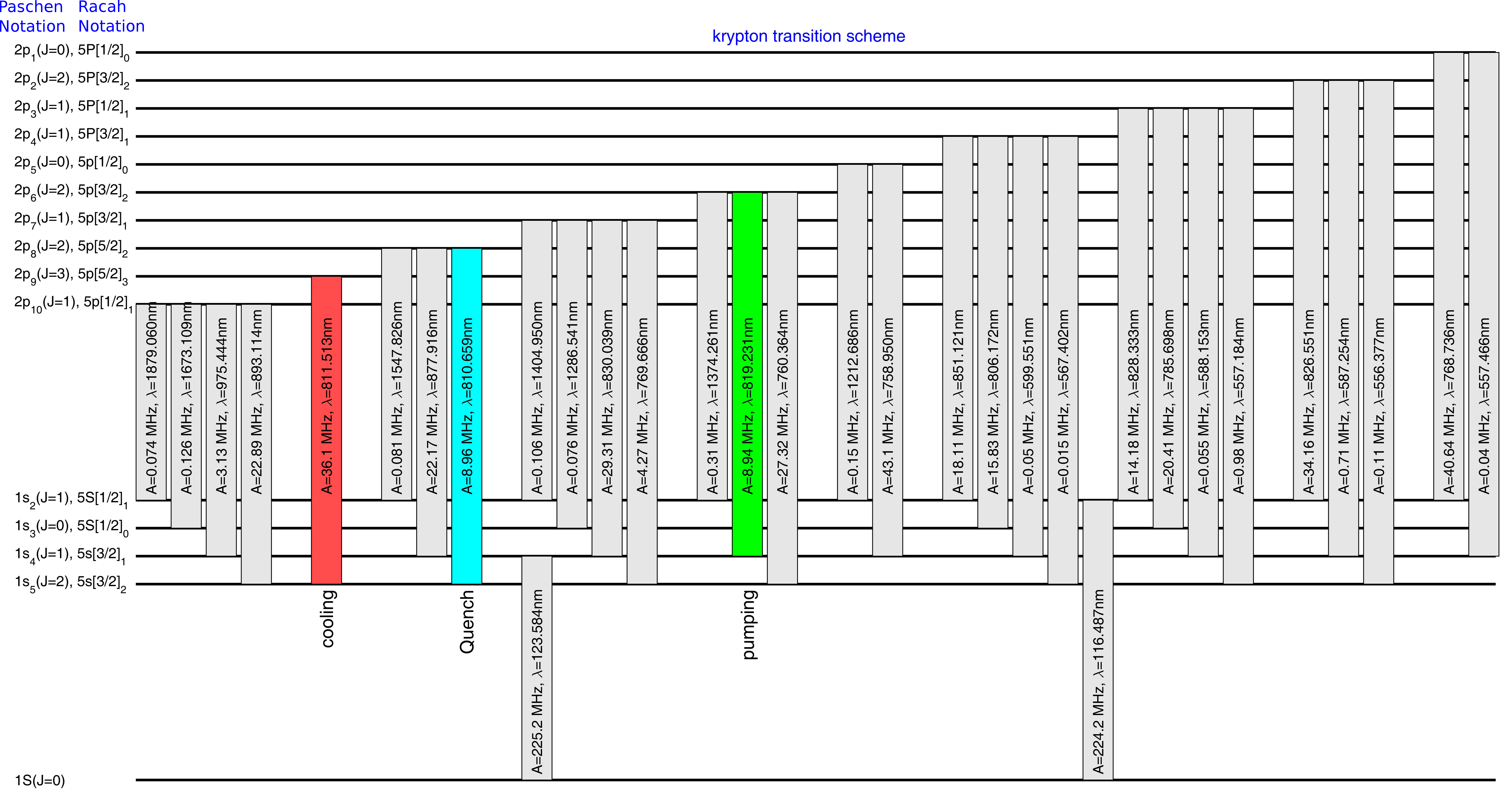}
			\caption{Krypton transition scheme calculated based on \cite{NIST_ASD}. States with capital letter in the Racah notation refer to $ j_{\text{core}}=3/2 $ while states with small letters refer to $ j_{\text{core}}=1/2 $.}
			\label{transition}
		\end{figure}

		To determine the most suitable transitions for optical pumping to the metastable state $ 1s_5 $, we have theoretically investigated all the $ 1s-2p $ transitions in krypton. The transfer efficiency for each transition has been calculated according to the derivation in Sec. \ref{theory_transfer_efficiency} and the results are compiled in Table \ref{tab:trans_frac}. For each $ 1s $ state we can thereby identify the transition with the highest transfer efficiency. Among these, we experimentally found the $ 1s_{4}-2p_{6} $ transition at \SI{819}{nm} to be the strongest one for optical pumping and therefore have chosen it for this work. A scheme of all $ 1s-2p $ transitions in krypton is illustrated in Fig. \ref{transition} with the levels in Racah as well as in Paschen notation.
		
		\begin{table}[H]
			\centering	
			\def\arraystretch{1.3} 
			\caption{Transfer efficiencies $ \widetilde{\rho}_{44}(t\to+\infty) $ for $ 1s-2p $ transitions in \Kr{84} calculated for a laser beam with 9-mm diameter and different powers $ P $. The transitions highlighted in bold are the ones with the highest transfer efficiency. The levels are provided in Paschen as well as in Racah notation.}
			\begin{tabular*}{0.618\hsize}{@{}@{\extracolsep{\fill}}ccccc@{}}
				\\ \hline\hline
				\multirow{2}*{Lower state}&\multirow{2}*{Upper state}&\multirow{2}*{$\lambda \ \SI{}{(nm)}$}&\multicolumn{2}{c}{Transfer efficiency $ \widetilde{\rho}_{44}(t\to+\infty) $} \\ \cline{4-5} &&&$P=\SI{0.5}{W}$&$ P\rightarrow +\infty \SI{}{W} $  \\ \hline
				\multirow{7}*{$1s_4$, $5s[3/2]_1$}&$2p_{10}$, $5p[1/2]_1$&975.44&0.069&0.092 \\
				&$2p_8$, $5p[5/2]_2$&877.92&0.036&0.038\\ 
				&$2p_7$, $5p[3/2]_1$&830.04&0.017&0.019\\ 
				&\boldmath{$2p_6$, $5p[3/2]_2$}&\textbf{819.23}&\textbf{0.093}&\textbf{0.108}  \\ 
				&$2p_4$, $5P[3/2]_1$&599.55&0.000&0.000\\ 
				&$2p_3$, $5P[1/2]_1$&588.15&0.000&0.004 \\ 
				&$2p_2$, $5P[3/2]_2$&587.25&0.000&0.000
				\\ \hline
				\multirow{4}*{$1s_3$, $5S[1/2]_0$}
				&\boldmath{$2p_{10}$, $5p[1/2]_1$}&\textbf{1673.11}&\textbf{0.877}&\textbf{0.877} \\ 
				&$2p_7$, $5p[3/2]_1$&1286.54&0.127&0.127\\ 
				&$2p_4$, $5P[3/2]_1$&806.17&0.001&0.001\\ 
				
				&$2p_3$, $5P[1/2]_1$&785.70&0.064&0.064
				\\ \hline
				\multirow{7}*{$1s_2$, $5s[1/2]_1$}&$2p_{10}$, $5p[1/2]_1$&1879.06&0.032&0.091\\
				
				&$2p_8$, $5p[5/2]_2$&1547.83&0.011&0.035 \\ 
				&$2p_7$, $5p[3/2]_1$&1404.95&0.003&0.017 \\
				
				&\boldmath{$2p_6$, $5p[3/2]_2$}&\textbf{1374.26}&\textbf{0.053}&\textbf{0.105} \\ 
				&$2p_4$, $5P[3/2]_1$&851.12&0.000&0.000\\ 
				&$2p_3$, $5P[1/2]_1$&828.33&0.003&0.004\\ 
				&$2p_2$, $5P[3/2]_2$&826.55&0.000&0.000\\  \hline\hline
			\end{tabular*}
			\label{tab:trans_frac}
		\end{table}
		
	\clearpage

	\section{ISOTOPE, HYPERFINE, AND TOTAL FREQUENCY SHIFTS FOR THE 819-nm TRANSITION IN ODD KRYPTON ISOTOPES}\label{A_B_kr_cal}	
	\begin{table}[H]
		\centering
		\def\arraystretch{1.3}
		\caption{Hyperfine coefficients $ A $ and $ B $ for the odd krypton isotopes and different levels.}
		\begin{tabular*}{0.309\hsize}{@{}@{\extracolsep{\fill}}cccc@{}}  
			\hline\hline
			Isotope                                   & State          & $ A $(MHz) & $ B $(MHz) \\ \hline
			\multirow{2}{*}{\Kr{81}} & $ 1s_4 $ & $ -193.2 $\hyperref[a]{$ ^{\text a} $} & $ -263.3 $\hyperref[a]{$ ^{\text a} $} \\  
			& $ 2p_6 $ & $ -130.5 $\hyperref[b]{$ ^{\text b} $} & $ -216 $\hyperref[b]{$ ^{\text b} $}   \\ \hline
			\multirow{2}{*}{\Kr{83}} & $ 1s_4 $ & $ -160.5 $\hyperref[c]{$ ^{\text c} $} & $ -105.9 $\hyperref[c]{$ ^{\text c} $} \\  
			& $ 2p_6 $ & $ -108.4 $\hyperref[b]{$ ^{\text b} $} & $ -86 $\hyperref[b]{$ ^{\text b} $}    \\ \hline
			\multirow{2}{*}{\Kr{85}} & $ 1s_4 $ & $ -166.2 $\hyperref[a]{$ ^{\text a} $} & $ -181.2 $\hyperref[a]{$ ^{\text a} $} \\ 
			& $ 2p_6 $ & $ -112.5 $\hyperref[b]{$ ^{\text b} $} & $ -151 $\hyperref[b]{$ ^{\text b} $}   \\ 
			\hline\hline
		\end{tabular*}
	    \label{tab:A_B}
	\end{table}
\begin{center}
	\footnotesize
	\label{a}{$ ^{\text a} $}Calculated in this work.\hspace{24pt}
	\label{b}{$ ^{\text b} $}Reference \cite{Cannon1993}.\hspace{24pt}
	\label{c}{$ ^{\text c} $}Reference \cite{Jackson1977}.	
\end{center}

	\begin{table}[h]
		\centering
		\def\arraystretch{1.3}
		\caption{Isotope, hyperfine and total frequency shifts for the 819-nm transition in odd krypton isotopes. The hyperfine shift is relative to the center of gravity of the fine-structure term and the isotope shift is relative to \Kr{84}.}
		\begin{tabular*}{0.618\hsize}{@{}@{\extracolsep{\fill}}cccccc@{}}  
			\\ \hline\hline
			Isotope               & \begin{tabular}[c]{@{}c@{}}Isotope shift\\ MHz\end{tabular}           & \begin{tabular}[c]{@{}c@{}}Lower state\\ $ 1s_4 $\end{tabular}              & \begin{tabular}[c]{@{}c@{}}Upper state\\ $ 2p_6 $\end{tabular}  & \begin{tabular}[c]{@{}c@{}}HFS shift\\ MHz\end{tabular} & \begin{tabular}[c]{@{}c@{}}Total shift\\ MHz\end{tabular} \\ \hline
			\multirow{9}{*}{\Kr{81}} & \multirow{9}{*}{$ -160.9 $} & \multirow{3}{*}{$ F=5/2 $}  & $ F=3/2 $       & 330.5     & 169.6       \\ 
			&                         &                         & $ F=5/2 $       & 100.7     & $ -60.2 $       \\ 
			&                         &                         & $ F=7/2 $       & $ -275.1 $    & $ -436 $        \\ \cline{3-6}
			&                         & \multirow{3}{*}{$ F=7/2 $}  & $ F=5/2 $       & 447.7     & 286.8       \\ 
			&                         &                         & $ F=7/2 $       & 71.9      & $ -89 $         \\ 
			&                         &                         & $ F=9/2 $       & $ -503.8 $    & $ -664.7 $      \\ \cline{3-6}
			&                         & \multirow{3}{*}{$ F=9/2 $}  & $ F=7/2 $       & 1195.2    & 1034.3      \\ 
			&                         &                         & $ F=9/2 $       & 619.5     & 458.6       \\ 
			&                         &                         & $ F=11/2 $      & $ -225.5 $    & $ -386.4 $      \\ \cline{1-6}
			\multirow{9}{*}{\Kr{83}} & \multirow{9}{*}{$ -44.8 $}  & \multirow{3}{*}{$ F=7/2 $}  & $ F=5/2 $       & 318.8     & 274         \\ 
			&                         &                         & $ F=7/2 $       & $ -23 $       & $ -67.8 $       \\ 
			&                         &                         & $ F=9/2 $       & $ -483.9 $    & $ -528.7 $      \\ \cline{3-6}
			&                         & \multirow{3}{*}{$ F=9/2 $}  & $ F=7/2 $       & 580.1     & 535.3       \\ 
			&                         &                         & $ F=9/2 $       & 119.2     & 74.4        \\ 
			&                         &                         & $ F=11/2 $      & $ -477 $      & $ -521.8 $      \\ \cline{3-6}
			&                         & \multirow{3}{*}{$ F=11/2 $} & $ F=9/2 $       & 1099      & 1054.2      \\ 
			&                         &                         & $ F=11/2 $      & 502.8     & 458         \\ 
			&                         &                         & $ F=13/2 $      & $ -248.4 $    & $ -293.2 $      \\ \cline{1-6}
			\multirow{9}{*}{\Kr{85}} & \multirow{9}{*}{52.9}   & \multirow{3}{*}{$ F=7/2 $}  & $ F=5/2 $       & 337.3     & 390.2       \\ 
			&                         &                         & $ F=7/2 $       & 9.6       & 62.5        \\ 
			&                         &                         & $ F=9/2 $       & $ -449.5 $    & $ -396.6 $      \\ \cline{3-6}
			&                         & \multirow{3}{*}{$ F=9/2 $}  & $ F=7/2 $       & 553.6     & 606.5       \\ 
			&                         &                         & $ F=9/2 $       & 94.5      & 147.4       \\ 
			&                         &                         & $ F=11/2 $      & $ -524.2 $    & $ -471.3 $      \\ \cline{3-6}
			&                         & \multirow{3}{*}{$ F=11/2 $} & $ F=9/2 $       & 1174.7    & 1227.6      \\ 
			&                         &                         & $ F=11/2 $      & 556       & 608.9       \\ 
			&                         &                         & $ F=13/2 $      & $ -257 $      & $ -204.1 $      \\ \hline\hline
		\end{tabular*}
		\label{tab:kr_cal}
	\end{table}

	\newpage
	\twocolumngrid
%
	

\begin{thebibliography}{30}%
	\makeatletter
	\providecommand \@ifxundefined [1]{%
		\@ifx{#1\undefined}
	}%
	\providecommand \@ifnum [1]{%
		\ifnum #1\expandafter \@firstoftwo
		\else \expandafter \@secondoftwo
		\fi
	}%
	\providecommand \@ifx [1]{%
		\ifx #1\expandafter \@firstoftwo
		\else \expandafter \@secondoftwo
		\fi
	}%
	\providecommand \natexlab [1]{#1}%
	\providecommand \enquote  [1]{``#1''}%
	\providecommand \bibnamefont  [1]{#1}%
	\providecommand \bibfnamefont [1]{#1}%
	\providecommand \citenamefont [1]{#1}%
	\providecommand \href@noop [0]{\@secondoftwo}%
	\providecommand \href [0]{\begingroup \@sanitize@url \@href}%
	\providecommand \@href[1]{\@@startlink{#1}\@@href}%
	\providecommand \@@href[1]{\endgroup#1\@@endlink}%
	\providecommand \@sanitize@url [0]{\catcode `\\12\catcode `\$12\catcode
		`\&12\catcode `\#12\catcode `\^12\catcode `\_12\catcode `\%12\relax}%
	\providecommand \@@startlink[1]{}%
	\providecommand \@@endlink[0]{}%
	\providecommand \url  [0]{\begingroup\@sanitize@url \@url }%
	\providecommand \@url [1]{\endgroup\@href {#1}{\urlprefix }}%
	\providecommand \urlprefix  [0]{URL }%
	\providecommand \Eprint [0]{\href }%
	\providecommand \doibase [0]{https://doi.org/}%
	\providecommand \selectlanguage [0]{\@gobble}%
	\providecommand \bibinfo  [0]{\@secondoftwo}%
	\providecommand \bibfield  [0]{\@secondoftwo}%
	\providecommand \translation [1]{[#1]}%
	\providecommand \BibitemOpen [0]{}%
	\providecommand \bibitemStop [0]{}%
	\providecommand \bibitemNoStop [0]{.\EOS\space}%
	\providecommand \EOS [0]{\spacefactor3000\relax}%
	\providecommand \BibitemShut  [1]{\csname bibitem#1\endcsname}%
	\let\auto@bib@innerbib\@empty
	\bibitem [{\citenamefont {Loosli}\ and\ \citenamefont
		{Oeschger}(1969)}]{Loosli1969}%
	\BibitemOpen
	\bibfield  {author} {\bibinfo {author} {\bibfnamefont {H.~H.}\ \bibnamefont
			{Loosli}}\ and\ \bibinfo {author} {\bibfnamefont {H.}~\bibnamefont
			{Oeschger}},\ }\href {https://doi.org/10.1016/0012-821X(69)90014-4}
	{\bibfield  {journal} {\bibinfo  {journal} {Earth and Planetary Science
				Letters}\ }\textbf {\bibinfo {volume} {7}},\ \bibinfo {pages} {67} (\bibinfo
		{year} {1969})}\BibitemShut {NoStop}%
	\bibitem [{\citenamefont {Lu}\ \emph {et~al.}(2014)\citenamefont {Lu},
		\citenamefont {Schlosser}, \citenamefont {Smethie}, \citenamefont {Sturchio},
		\citenamefont {Fischer}, \citenamefont {Kennedy}, \citenamefont {Purtschert},
		\citenamefont {Severinghaus}, \citenamefont {Solomon}, \citenamefont
		{Tanhua},\ and\ \citenamefont {Yokochi}}]{Lu2014}%
	\BibitemOpen
	\bibfield  {author} {\bibinfo {author} {\bibfnamefont {Z.~T.}\ \bibnamefont
			{Lu}}, \bibinfo {author} {\bibfnamefont {P.}~\bibnamefont {Schlosser}},
		\bibinfo {author} {\bibfnamefont {W.~M.}\ \bibnamefont {Smethie}}, \bibinfo
		{author} {\bibfnamefont {N.~C.}\ \bibnamefont {Sturchio}}, \bibinfo {author}
		{\bibfnamefont {T.~P.}\ \bibnamefont {Fischer}}, \bibinfo {author}
		{\bibfnamefont {B.~M.}\ \bibnamefont {Kennedy}}, \bibinfo {author}
		{\bibfnamefont {R.}~\bibnamefont {Purtschert}}, \bibinfo {author}
		{\bibfnamefont {J.~P.}\ \bibnamefont {Severinghaus}}, \bibinfo {author}
		{\bibfnamefont {D.~K.}\ \bibnamefont {Solomon}}, \bibinfo {author}
		{\bibfnamefont {T.}~\bibnamefont {Tanhua}},\ and\ \bibinfo {author}
		{\bibfnamefont {R.}~\bibnamefont {Yokochi}},\ }\href
	{https://doi.org/10.1016/j.earscirev.2013.09.002} {\bibfield  {journal}
		{\bibinfo  {journal} {Earth-Science Reviews}\ }\textbf {\bibinfo {volume}
			{138}},\ \bibinfo {pages} {196} (\bibinfo {year} {2014})}\BibitemShut
	{NoStop}%
	\bibitem [{\citenamefont {Chen}\ \emph {et~al.}(1999)\citenamefont {Chen},
		\citenamefont {Li}, \citenamefont {Bailey}, \citenamefont {O'Connor},
		\citenamefont {Young},\ and\ \citenamefont {Lu}}]{Chen1999}%
	\BibitemOpen
	\bibfield  {author} {\bibinfo {author} {\bibfnamefont {C.~Y.}\ \bibnamefont
			{Chen}}, \bibinfo {author} {\bibfnamefont {Y.~M.}\ \bibnamefont {Li}},
		\bibinfo {author} {\bibfnamefont {K.}~\bibnamefont {Bailey}}, \bibinfo
		{author} {\bibfnamefont {T.~P.}\ \bibnamefont {O'Connor}}, \bibinfo {author}
		{\bibfnamefont {L.}~\bibnamefont {Young}},\ and\ \bibinfo {author}
		{\bibfnamefont {Z.-T.}\ \bibnamefont {Lu}},\ }\href
	{https://doi.org/10.1126/science.286.5442.1139} {\bibfield  {journal}
		{\bibinfo  {journal} {Science}\ }\textbf {\bibinfo {volume} {286}},\ \bibinfo
		{pages} {1139} (\bibinfo {year} {1999})}\BibitemShut {NoStop}%
	\bibitem [{\citenamefont {Jiang}\ \emph {et~al.}(2012)\citenamefont {Jiang},
		\citenamefont {Bailey}, \citenamefont {Lu}, \citenamefont {Mueller},
		\citenamefont {O'Connor}, \citenamefont {Cheng}, \citenamefont {Hu},
		\citenamefont {Purtschert}, \citenamefont {Sturchio}, \citenamefont {Sun},
		\citenamefont {Williams},\ and\ \citenamefont {Yang}}]{Jiang2012}%
	\BibitemOpen
	\bibfield  {author} {\bibinfo {author} {\bibfnamefont {W.}~\bibnamefont
			{Jiang}}, \bibinfo {author} {\bibfnamefont {K.}~\bibnamefont {Bailey}},
		\bibinfo {author} {\bibfnamefont {Z.~T.}\ \bibnamefont {Lu}}, \bibinfo
		{author} {\bibfnamefont {P.}~\bibnamefont {Mueller}}, \bibinfo {author}
		{\bibfnamefont {T.~P.}\ \bibnamefont {O'Connor}}, \bibinfo {author}
		{\bibfnamefont {C.~F.}\ \bibnamefont {Cheng}}, \bibinfo {author}
		{\bibfnamefont {S.~M.}\ \bibnamefont {Hu}}, \bibinfo {author} {\bibfnamefont
			{R.}~\bibnamefont {Purtschert}}, \bibinfo {author} {\bibfnamefont {N.~C.}\
			\bibnamefont {Sturchio}}, \bibinfo {author} {\bibfnamefont {Y.~R.}\
			\bibnamefont {Sun}}, \bibinfo {author} {\bibfnamefont {W.~D.}\ \bibnamefont
			{Williams}},\ and\ \bibinfo {author} {\bibfnamefont {G.~M.}\ \bibnamefont
			{Yang}},\ }\href {https://doi.org/10.1016/j.gca.2012.05.019} {\bibfield
		{journal} {\bibinfo  {journal} {Geochimica et Cosmochimica Acta}\ }\textbf
		{\bibinfo {volume} {91}},\ \bibinfo {pages} {1} (\bibinfo {year}
		{2012})}\BibitemShut {NoStop}%
	\bibitem [{\citenamefont {Tian}\ \emph {et~al.}(2019)\citenamefont {Tian},
		\citenamefont {Ritterbusch}, \citenamefont {Gu}, \citenamefont {Hu},
		\citenamefont {Jiang}, \citenamefont {Lu}, \citenamefont {Wang},\ and\
		\citenamefont {Yang}}]{Tian2019}%
	\BibitemOpen
	\bibfield  {author} {\bibinfo {author} {\bibfnamefont {L.}~\bibnamefont
			{Tian}}, \bibinfo {author} {\bibfnamefont {F.}~\bibnamefont {Ritterbusch}},
		\bibinfo {author} {\bibfnamefont {J.-Q.}\ \bibnamefont {Gu}}, \bibinfo
		{author} {\bibfnamefont {S.-M.}\ \bibnamefont {Hu}}, \bibinfo {author}
		{\bibfnamefont {W.}~\bibnamefont {Jiang}}, \bibinfo {author} {\bibfnamefont
			{Z.-T.}\ \bibnamefont {Lu}}, \bibinfo {author} {\bibfnamefont
			{D.}~\bibnamefont {Wang}},\ and\ \bibinfo {author} {\bibfnamefont {G.-M.}\
			\bibnamefont {Yang}},\ }\href {https://doi.org/10.1029/2019GL082464}
	{\bibfield  {journal} {\bibinfo  {journal} {Geophysical Research Letters}\
		}\textbf {\bibinfo {volume} {46}},\ \bibinfo {pages} {6636} (\bibinfo {year}
		{2019})}\BibitemShut {NoStop}%
	\bibitem [{\citenamefont {Schaefer}\ \emph {et~al.}(2016)\citenamefont
		{Schaefer}, \citenamefont {Finkel}, \citenamefont {Balco}, \citenamefont
		{Alley}, \citenamefont {Caffee}, \citenamefont {Briner}, \citenamefont
		{Young}, \citenamefont {Gow},\ and\ \citenamefont {Schwartz}}]{Schaefer2016}%
	\BibitemOpen
	\bibfield  {author} {\bibinfo {author} {\bibfnamefont {J.~M.}\ \bibnamefont
			{Schaefer}}, \bibinfo {author} {\bibfnamefont {R.~C.}\ \bibnamefont
			{Finkel}}, \bibinfo {author} {\bibfnamefont {G.}~\bibnamefont {Balco}},
		\bibinfo {author} {\bibfnamefont {R.~B.}\ \bibnamefont {Alley}}, \bibinfo
		{author} {\bibfnamefont {M.~W.}\ \bibnamefont {Caffee}}, \bibinfo {author}
		{\bibfnamefont {J.~P.}\ \bibnamefont {Briner}}, \bibinfo {author}
		{\bibfnamefont {N.~E.}\ \bibnamefont {Young}}, \bibinfo {author}
		{\bibfnamefont {A.~J.}\ \bibnamefont {Gow}},\ and\ \bibinfo {author}
		{\bibfnamefont {R.}~\bibnamefont {Schwartz}},\ }\href
	{https://doi.org/10.1038/nature20146} {\bibfield  {journal} {\bibinfo
			{journal} {Nature}\ }\textbf {\bibinfo {volume} {540}},\ \bibinfo {pages}
		{252} (\bibinfo {year} {2016})}\BibitemShut {NoStop}%
	\bibitem [{\citenamefont {Yau}\ \emph {et~al.}(2016)\citenamefont {Yau},
		\citenamefont {Bender}, \citenamefont {Blunier},\ and\ \citenamefont
		{Jouzel}}]{Yau2016}%
	\BibitemOpen
	\bibfield  {author} {\bibinfo {author} {\bibfnamefont {A.~M.}\ \bibnamefont
			{Yau}}, \bibinfo {author} {\bibfnamefont {M.~L.}\ \bibnamefont {Bender}},
		\bibinfo {author} {\bibfnamefont {T.}~\bibnamefont {Blunier}},\ and\ \bibinfo
		{author} {\bibfnamefont {J.}~\bibnamefont {Jouzel}},\ }\href
	{https://doi.org/10.1016/j.epsl.2016.06.053} {\bibfield  {journal} {\bibinfo
			{journal} {Earth and Planetary Science Letters}\ }\textbf {\bibinfo {volume}
			{451}},\ \bibinfo {pages} {1} (\bibinfo {year} {2016})}\BibitemShut {NoStop}%
	\bibitem [{\citenamefont {Severinghaus}\ \emph {et~al.}(2010)\citenamefont
		{Severinghaus}, \citenamefont {Wolff},\ and\ \citenamefont
		{Brook}}]{Severinghaus2010}%
	\BibitemOpen
	\bibfield  {author} {\bibinfo {author} {\bibfnamefont {J.}~\bibnamefont
			{Severinghaus}}, \bibinfo {author} {\bibfnamefont {E.~W.}\ \bibnamefont
			{Wolff}},\ and\ \bibinfo {author} {\bibfnamefont {E.~J.}\ \bibnamefont
			{Brook}},\ }\href {https://doi.org/10.1029/2010EO400001} {\bibfield
		{journal} {\bibinfo  {journal} {Eos, Transactions American Geophysical
				Union}\ }\textbf {\bibinfo {volume} {91}},\ \bibinfo {pages} {357} (\bibinfo
		{year} {2010})}\BibitemShut {NoStop}%
	\bibitem [{\citenamefont {Paschen}(1919)}]{Paschen1919}%
	\BibitemOpen
	\bibfield  {author} {\bibinfo {author} {\bibfnamefont {F.}~\bibnamefont
			{Paschen}},\ }\href {https://doi.org/10.1002/andp.19193652102} {\bibfield
		{journal} {\bibinfo  {journal} {Annalen der Physik}\ }\textbf {\bibinfo
			{volume} {365}},\ \bibinfo {pages} {405} (\bibinfo {year}
		{1919})}\BibitemShut {NoStop}%
	\bibitem [{\citenamefont {Racah}(1942)}]{Racah1942}%
	\BibitemOpen
	\bibfield  {author} {\bibinfo {author} {\bibfnamefont {G.}~\bibnamefont
			{Racah}},\ }\href {https://doi.org/10.1103/PhysRev.61.537} {\bibfield
		{journal} {\bibinfo  {journal} {Physical Review}\ }\textbf {\bibinfo {volume}
			{61}},\ \bibinfo {pages} {537} (\bibinfo {year} {1942})}\BibitemShut
	{NoStop}%
	\bibitem [{\citenamefont {Hans}(2014)}]{Hans2014}%
	\BibitemOpen
	\bibfield  {author} {\bibinfo {author} {\bibfnamefont {M.}~\bibnamefont
			{Hans}},\ }\href@noop {} {\bibfield  {journal} {\bibinfo  {journal} {Bachelor
				thesis, Heidelberg University}\ } (\bibinfo {year} {2014})}\BibitemShut
	{NoStop}%
	\bibitem [{\citenamefont {Fr{\"o}lian}(2015)}]{Frolian2015}%
	\BibitemOpen
	\bibfield  {author} {\bibinfo {author} {\bibfnamefont {A.}~\bibnamefont
			{Fr{\"o}lian}},\ }\href@noop {} {\bibfield  {journal} {\bibinfo  {journal}
			{Bachelor thesis, Heidelberg University}\ } (\bibinfo {year}
		{2015})}\BibitemShut {NoStop}%
	\bibitem [{\citenamefont {Hickman}\ \emph {et~al.}(2016)\citenamefont
		{Hickman}, \citenamefont {Franson},\ and\ \citenamefont
		{Pittman}}]{Hickman2016}%
	\BibitemOpen
	\bibfield  {author} {\bibinfo {author} {\bibfnamefont {G.~T.}\ \bibnamefont
			{Hickman}}, \bibinfo {author} {\bibfnamefont {J.~D.}\ \bibnamefont
			{Franson}},\ and\ \bibinfo {author} {\bibfnamefont {T.~B.}\ \bibnamefont
			{Pittman}},\ }\href {https://doi.org/10.1364/OL.41.004372} {\bibfield
		{journal} {\bibinfo  {journal} {Optics Letters}\ }\textbf {\bibinfo {volume}
			{41}},\ \bibinfo {pages} {4372} (\bibinfo {year} {2016})}\BibitemShut
	{NoStop}%
	\bibitem [{\citenamefont {Steck}(2001)}]{Steck2001}%
	\BibitemOpen
	\bibfield  {author} {\bibinfo {author} {\bibfnamefont {D.~A.}\ \bibnamefont
			{Steck}},\ }\href@noop {} {\bibinfo {title} {Rubidium 87 {{D}} line data}}
	(\bibinfo {year} {2001})\BibitemShut {NoStop}%
	\bibitem [{\citenamefont {Armstrong}(1971)}]{Armstrong1971}%
	\BibitemOpen
	\bibfield  {author} {\bibinfo {author} {\bibfnamefont {L.}~\bibnamefont
			{Armstrong}},\ }\href@noop {} {\emph {\bibinfo {title} {Theory of the
				Hyperfine Structure of Free Atoms}}}\ (\bibinfo  {publisher}
	{{Wiley-Interscience}},\ \bibinfo {address} {{New York}},\ \bibinfo {year}
	{1971})\ \bibinfo {note} {oCLC: 639161041}\BibitemShut {NoStop}%
	\bibitem [{\citenamefont {Cannon}(1993)}]{Cannon1993}%
	\BibitemOpen
	\bibfield  {author} {\bibinfo {author} {\bibfnamefont {B.~D.}\ \bibnamefont
			{Cannon}},\ }\href {https://doi.org/10.1103/PhysRevA.47.1148} {\bibfield
		{journal} {\bibinfo  {journal} {Physical Review A}\ }\textbf {\bibinfo
			{volume} {47}},\ \bibinfo {pages} {1148} (\bibinfo {year}
		{1993})}\BibitemShut {NoStop}%
	\bibitem [{\citenamefont {Jackson}(1977)}]{Jackson1977}%
	\BibitemOpen
	\bibfield  {author} {\bibinfo {author} {\bibfnamefont {D.~A.}\ \bibnamefont
			{Jackson}},\ }\href {https://doi.org/10.1364/JOSA.67.001638} {\bibfield
		{journal} {\bibinfo  {journal} {JOSA}\ }\textbf {\bibinfo {volume} {67}},\
		\bibinfo {pages} {1638} (\bibinfo {year} {1977})}\BibitemShut {NoStop}%
	\bibitem [{\citenamefont {Jackson}(1979)}]{Jackson1979}%
	\BibitemOpen
	\bibfield  {author} {\bibinfo {author} {\bibfnamefont {D.~A.}\ \bibnamefont
			{Jackson}},\ }\href {https://doi.org/10.1364/JOSA.69.000503} {\bibfield
		{journal} {\bibinfo  {journal} {JOSA}\ }\textbf {\bibinfo {volume} {69}},\
		\bibinfo {pages} {503} (\bibinfo {year} {1979})}\BibitemShut {NoStop}%
	\bibitem [{\citenamefont {Heilig}\ and\ \citenamefont
		{Steudel}(1974)}]{Heilig1974}%
	\BibitemOpen
	\bibfield  {author} {\bibinfo {author} {\bibfnamefont {K.}~\bibnamefont
			{Heilig}}\ and\ \bibinfo {author} {\bibfnamefont {A.}~\bibnamefont
			{Steudel}},\ }\href {https://doi.org/10.1016/S0092-640X(74)80006-9}
	{\bibfield  {journal} {\bibinfo  {journal} {Atomic Data and Nuclear Data
				Tables}\ }\bibinfo {series} {Nuclear {{Charge}} and {{Moment
					Distributions}}},\ \textbf {\bibinfo {volume} {14}},\ \bibinfo {pages} {613}
		(\bibinfo {year} {1974})}\BibitemShut {NoStop}%
	\bibitem [{\citenamefont {Keim}\ \emph {et~al.}(1995)\citenamefont {Keim},
		\citenamefont {Arnold}, \citenamefont {Borchers}, \citenamefont {Georg},
		\citenamefont {Klein}, \citenamefont {Neugart}, \citenamefont {Vermeeren},
		\citenamefont {Silverans},\ and\ \citenamefont {Lievens}}]{Keim1995}%
	\BibitemOpen
	\bibfield  {author} {\bibinfo {author} {\bibfnamefont {M.}~\bibnamefont
			{Keim}}, \bibinfo {author} {\bibfnamefont {E.}~\bibnamefont {Arnold}},
		\bibinfo {author} {\bibfnamefont {W.}~\bibnamefont {Borchers}}, \bibinfo
		{author} {\bibfnamefont {U.}~\bibnamefont {Georg}}, \bibinfo {author}
		{\bibfnamefont {A.}~\bibnamefont {Klein}}, \bibinfo {author} {\bibfnamefont
			{R.}~\bibnamefont {Neugart}}, \bibinfo {author} {\bibfnamefont
			{L.}~\bibnamefont {Vermeeren}}, \bibinfo {author} {\bibfnamefont {R.~E.}\
			\bibnamefont {Silverans}},\ and\ \bibinfo {author} {\bibfnamefont
			{P.}~\bibnamefont {Lievens}},\ }\href
	{https://doi.org/10.1016/0375-9474(94)00786-M} {\bibfield  {journal}
		{\bibinfo  {journal} {Nuclear Physics A}\ }\textbf {\bibinfo {volume}
			{586}},\ \bibinfo {pages} {219} (\bibinfo {year} {1995})}\BibitemShut
	{NoStop}%
	\bibitem [{\citenamefont {Cheng}\ \emph {et~al.}(2013)\citenamefont {Cheng},
		\citenamefont {Yang}, \citenamefont {Jiang}, \citenamefont {Sun},
		\citenamefont {Tu},\ and\ \citenamefont {Hu}}]{Cheng2013}%
	\BibitemOpen
	\bibfield  {author} {\bibinfo {author} {\bibfnamefont {C.~F.}\ \bibnamefont
			{Cheng}}, \bibinfo {author} {\bibfnamefont {G.~M.}\ \bibnamefont {Yang}},
		\bibinfo {author} {\bibfnamefont {W.}~\bibnamefont {Jiang}}, \bibinfo
		{author} {\bibfnamefont {Y.~R.}\ \bibnamefont {Sun}}, \bibinfo {author}
		{\bibfnamefont {L.~Y.}\ \bibnamefont {Tu}},\ and\ \bibinfo {author}
		{\bibfnamefont {S.~M.}\ \bibnamefont {Hu}},\ }\href
	{https://doi.org/10.1364/OL.38.000031} {\bibfield  {journal} {\bibinfo
			{journal} {Optics Letters}\ }\textbf {\bibinfo {volume} {38}},\ \bibinfo
		{pages} {31} (\bibinfo {year} {2013})}\BibitemShut {NoStop}%
	\bibitem [{\citenamefont {Ma}\ and\ \citenamefont {Hall}(1990)}]{Ma1990}%
	\BibitemOpen
	\bibfield  {author} {\bibinfo {author} {\bibfnamefont {L.}~\bibnamefont
			{Ma}}\ and\ \bibinfo {author} {\bibfnamefont {J.~L.}\ \bibnamefont {Hall}},\
	}\href {https://doi.org/10.1109/3.62120} {\bibfield  {journal} {\bibinfo
			{journal} {IEEE Journal of Quantum Electronics}\ }\textbf {\bibinfo {volume}
			{26}},\ \bibinfo {pages} {2006} (\bibinfo {year} {1990})}\BibitemShut
	{NoStop}%
	\bibitem [{\citenamefont {Zhao}\ \emph {et~al.}(1998)\citenamefont {Zhao},
		\citenamefont {Simsarian}, \citenamefont {Orozco},\ and\ \citenamefont
		{Sprouse}}]{Zhao1998}%
	\BibitemOpen
	\bibfield  {author} {\bibinfo {author} {\bibfnamefont {W.~Z.}\ \bibnamefont
			{Zhao}}, \bibinfo {author} {\bibfnamefont {J.~E.}\ \bibnamefont {Simsarian}},
		\bibinfo {author} {\bibfnamefont {L.~A.}\ \bibnamefont {Orozco}},\ and\
		\bibinfo {author} {\bibfnamefont {G.~D.}\ \bibnamefont {Sprouse}},\ }\href
	{https://doi.org/10.1063/1.1149171} {\bibfield  {journal} {\bibinfo
			{journal} {Review of Scientific Instruments}\ }\textbf {\bibinfo {volume}
			{69}},\ \bibinfo {pages} {3737} (\bibinfo {year} {1998})}\BibitemShut
	{NoStop}%
	\bibitem [{\citenamefont {Subhankar}\ \emph {et~al.}(2019)\citenamefont
		{Subhankar}, \citenamefont {Restelli}, \citenamefont {Wang}, \citenamefont
		{Rolston},\ and\ \citenamefont {Porto}}]{Subhankar2019}%
	\BibitemOpen
	\bibfield  {author} {\bibinfo {author} {\bibfnamefont {S.}~\bibnamefont
			{Subhankar}}, \bibinfo {author} {\bibfnamefont {A.}~\bibnamefont {Restelli}},
		\bibinfo {author} {\bibfnamefont {Y.}~\bibnamefont {Wang}}, \bibinfo {author}
		{\bibfnamefont {S.~L.}\ \bibnamefont {Rolston}},\ and\ \bibinfo {author}
		{\bibfnamefont {J.~V.}\ \bibnamefont {Porto}},\ }\href
	{https://doi.org/10.1063/1.5067266} {\bibfield  {journal} {\bibinfo
			{journal} {Review of Scientific Instruments}\ }\textbf {\bibinfo {volume}
			{90}},\ \bibinfo {pages} {043115} (\bibinfo {year} {2019})}\BibitemShut
	{NoStop}%
	\bibitem [{\citenamefont {Axner}\ \emph {et~al.}(2004)\citenamefont {Axner},
		\citenamefont {Gustafsson}, \citenamefont {Omenetto},\ and\ \citenamefont
		{Winefordner}}]{Axner2004}%
	\BibitemOpen
	\bibfield  {author} {\bibinfo {author} {\bibfnamefont {O.}~\bibnamefont
			{Axner}}, \bibinfo {author} {\bibfnamefont {J.}~\bibnamefont {Gustafsson}},
		\bibinfo {author} {\bibfnamefont {N.}~\bibnamefont {Omenetto}},\ and\
		\bibinfo {author} {\bibfnamefont {J.~D.}\ \bibnamefont {Winefordner}},\
	}\href {https://doi.org/10.1016/j.sab.2003.10.002} {\bibfield  {journal}
		{\bibinfo  {journal} {Spectrochimica Acta Part B: Atomic Spectroscopy}\
		}\textbf {\bibinfo {volume} {59}},\ \bibinfo {pages} {1} (\bibinfo {year}
		{2004})}\BibitemShut {NoStop}%
	\bibitem [{\citenamefont {Young}\ \emph {et~al.}(2002)\citenamefont {Young},
		\citenamefont {Yang},\ and\ \citenamefont {Dunford}}]{Young2002}%
	\BibitemOpen
	\bibfield  {author} {\bibinfo {author} {\bibfnamefont {L.}~\bibnamefont
			{Young}}, \bibinfo {author} {\bibfnamefont {D.}~\bibnamefont {Yang}},\ and\
		\bibinfo {author} {\bibfnamefont {R.~W.}\ \bibnamefont {Dunford}},\ }\href
	{https://doi.org/10.1088/0953-4075/35/13/311} {\bibfield  {journal} {\bibinfo
			{journal} {Journal of Physics B: Atomic, Molecular and Optical Physics}\
		}\textbf {\bibinfo {volume} {35}},\ \bibinfo {pages} {2985} (\bibinfo {year}
		{2002})}\BibitemShut {NoStop}%
	\bibitem [{\citenamefont {Ding}\ \emph {et~al.}(2007)\citenamefont {Ding},
		\citenamefont {Hu}, \citenamefont {Bailey}, \citenamefont {Davis},
		\citenamefont {Dunford}, \citenamefont {Lu}, \citenamefont {O'Connor},\ and\
		\citenamefont {Young}}]{Ding2007}%
	\BibitemOpen
	\bibfield  {author} {\bibinfo {author} {\bibfnamefont {Y.}~\bibnamefont
			{Ding}}, \bibinfo {author} {\bibfnamefont {S.-M.}\ \bibnamefont {Hu}},
		\bibinfo {author} {\bibfnamefont {K.}~\bibnamefont {Bailey}}, \bibinfo
		{author} {\bibfnamefont {A.~M.}\ \bibnamefont {Davis}}, \bibinfo {author}
		{\bibfnamefont {R.~W.}\ \bibnamefont {Dunford}}, \bibinfo {author}
		{\bibfnamefont {Z.-T.}\ \bibnamefont {Lu}}, \bibinfo {author} {\bibfnamefont
			{T.~P.}\ \bibnamefont {O'Connor}},\ and\ \bibinfo {author} {\bibfnamefont
			{L.}~\bibnamefont {Young}},\ }\href {https://doi.org/10.1063/1.2437193}
	{\bibfield  {journal} {\bibinfo  {journal} {Review of Scientific
				Instruments}\ }\textbf {\bibinfo {volume} {78}},\ \bibinfo {pages} {023103}
		(\bibinfo {year} {2007})}\BibitemShut {NoStop}%
	\bibitem [{\citenamefont {Daerr}\ \emph {et~al.}(2011)\citenamefont {Daerr},
		\citenamefont {Kohler}, \citenamefont {Sahling}, \citenamefont {Tippenhauer},
		\citenamefont {{Arabi-Hashemi}}, \citenamefont {Becker}, \citenamefont
		{Sengstock},\ and\ \citenamefont {Kalinowski}}]{Daerr2011}%
	\BibitemOpen
	\bibfield  {author} {\bibinfo {author} {\bibfnamefont {H.}~\bibnamefont
			{Daerr}}, \bibinfo {author} {\bibfnamefont {M.}~\bibnamefont {Kohler}},
		\bibinfo {author} {\bibfnamefont {P.}~\bibnamefont {Sahling}}, \bibinfo
		{author} {\bibfnamefont {S.}~\bibnamefont {Tippenhauer}}, \bibinfo {author}
		{\bibfnamefont {A.}~\bibnamefont {{Arabi-Hashemi}}}, \bibinfo {author}
		{\bibfnamefont {C.}~\bibnamefont {Becker}}, \bibinfo {author} {\bibfnamefont
			{K.}~\bibnamefont {Sengstock}},\ and\ \bibinfo {author} {\bibfnamefont
			{M.~B.}\ \bibnamefont {Kalinowski}},\ }\href
	{https://doi.org/10.1063/1.3610465} {\bibfield  {journal} {\bibinfo
			{journal} {Review of Scientific Instruments}\ }\textbf {\bibinfo {volume}
			{82}},\ \bibinfo {pages} {073106} (\bibinfo {year} {2011})}\BibitemShut
	{NoStop}%
	\bibitem [{\citenamefont {Kohler}\ \emph {et~al.}(2014)\citenamefont {Kohler},
		\citenamefont {Daerr}, \citenamefont {Sahling}, \citenamefont {Sieveke},
		\citenamefont {Jerschabek}, \citenamefont {Kalinowski}, \citenamefont
		{Becker},\ and\ \citenamefont {Sengstock}}]{Kohler2014}%
	\BibitemOpen
	\bibfield  {author} {\bibinfo {author} {\bibfnamefont {M.}~\bibnamefont
			{Kohler}}, \bibinfo {author} {\bibfnamefont {H.}~\bibnamefont {Daerr}},
		\bibinfo {author} {\bibfnamefont {P.}~\bibnamefont {Sahling}}, \bibinfo
		{author} {\bibfnamefont {C.}~\bibnamefont {Sieveke}}, \bibinfo {author}
		{\bibfnamefont {N.}~\bibnamefont {Jerschabek}}, \bibinfo {author}
		{\bibfnamefont {M.~B.}\ \bibnamefont {Kalinowski}}, \bibinfo {author}
		{\bibfnamefont {C.}~\bibnamefont {Becker}},\ and\ \bibinfo {author}
		{\bibfnamefont {K.}~\bibnamefont {Sengstock}},\ }\href
	{https://doi.org/10.1209/0295-5075/108/13001} {\bibfield  {journal} {\bibinfo
			{journal} {EPL (Europhysics Letters)}\ }\textbf {\bibinfo {volume} {108}},\
		\bibinfo {pages} {13001} (\bibinfo {year} {2014})}\BibitemShut {NoStop}%
	\bibitem [{\citenamefont {Kramida}\ \emph {et~al.}(2019)\citenamefont
		{Kramida}, \citenamefont {{Yu. Ralchenko}}, \citenamefont {Reader},\ and\
		\citenamefont {{NIST ASD Team}}}]{NIST_ASD}%
	\BibitemOpen
	\bibfield  {author} {\bibinfo {author} {\bibfnamefont {A.}~\bibnamefont
			{Kramida}}, \bibinfo {author} {\bibnamefont {{Yu. Ralchenko}}}, \bibinfo
		{author} {\bibfnamefont {J.}~\bibnamefont {Reader}},\ and\ \bibinfo {author}
		{\bibnamefont {{NIST ASD Team}}},\ }\href
	{https://doi.org/https://doi.org/10.18434/T4W30F} {}\bibinfo {howpublished}
	{{NIST Atomic Spectra Database (version 5.7.1), [Online]. Available:
			{\tt{https://physics.nist.gov/asd}} [Apr 14 2020]. National Institute of
			Standards and Technology, Gaithersburg, MD.} DOI:
		https://doi.org/10.18434/T4W30F} (\bibinfo {year} {2019})\BibitemShut
	{NoStop}%
\end{thebibliography}

\end{document}